\documentstyle[twoside,fleqn,espcrc2]{article}

    \newcommand{\cM}{{\cal{M}}}

\newcommand{\AmS}{{\protect\the\textfont2
  A\kern-.1667em\lower.5ex\hbox{M}\kern-.125emS}}

\title{The Quenched Approximation in Health and in Sickness
       \thanks{Presented by M. Golterman}}

\author{Claude Bernard\address{Department of Physics,
        Washington University\\
        St. Louis, MO 63130-4899, USA}%
        and
        Maarten Golterman}

\begin{document}

\begin{abstract}
We present results for physical quantities computed in quenched chiral
perturbation theory and compare them with the corresponding
unquenched expressions.
We also point out an apparent theoretical problem of
the quenched approximation.
\end{abstract}

\maketitle

\section{Introduction}
We have recently proposed a systematic lagrangian approach to chiral
perturbation theory (ChPT)
for the quenched approximation of QCD \cite{BerGolJap,BerGolPRD}.
This approach is based on the observation that QCD can be
quenched by adding a number of scalar (ghost)
quarks to the QCD lagrangian with
exactly the same quantum numbers and masses
as the physical, fermionic quarks, so that
the combined determinants cancel exactly \cite{Morel}.
This
defines a lagrangian
formalism for QCD in the quenched approximation.
The extended lagrangian possesses a graded $U(3|3)_L\otimes U(3|3)_R$ chiral
symmetry.

One can then develop ChPT for this larger symmetry
group in the usual way, assuming the existence of a Goldstone meson multiplet
which corresponds to the breakdown of this group to its diagonal subgroup.
This multiplet does not only contain the usual Goldstone mesons, but also
fermionic mesons which are bound states of a physical quark and a ghost
anti-quark or of a physical anti-quark and a ghost quark,
and ghost mesons which are ghost quark anti-quark
bound states.
One has to take into account the graded nature of the
symmetry group, by replacing traces and
determinants by super-traces and super-determinants in the construction of
the appropriate invariants.

\setcounter{footnote}{0}
In fact, the symmetry gets broken down to the semi-direct product group
[$SU(3|3)_L\otimes SU(3|3)_R]{\bigcirc\kern -0.25cm s\;}
U(1)$ by the anomaly,\footnote{The semi-direct product was incorrectly
identified as a direct product in refs. \cite{BerGolJap,BerGolPRD}.}
and correspondingly there exists a ``super-$\eta'$",
$\Phi_0$, which is invariant under this reduced symmetry group.  $\Phi_0$
is a linear combination of the $\eta'$ and its ghost partner,
${\tilde\eta}'$:
\[
\Phi_0=str\:\Phi=\frac{1}{\sqrt{2}}(\eta'-{\tilde\eta}'),
\]
where {\it str} denotes the super-trace and $\Phi$ is a graded hermitian
$6\times 6$
matrix describing the Goldstone meson multiplet (for details, see ref.
\cite{BerGolPRD}).

All these considerations lead to an $O(p^2)$ chiral lagrangian (in a notation
similar to the notation used in ref. \cite{GasLeut})
\begin{eqnarray*}
\lefteqn{L = -V_0(\Phi_0) +  V_1(\Phi_0)str(\partial_\mu \Sigma
\partial^\mu \Sigma^\dagger)} \\
& + V_2(\Phi_0)str(\cM \Sigma +\cM \Sigma^\dagger)
+ V_5(\Phi_0) (\partial_\mu\Phi_0)^2,
\end{eqnarray*}
where the functions $V_i$ can be chosen to be real and even,
$\Sigma=\exp(2 i \Phi/f)$, and
\begin{eqnarray*}
\cM = \left( \begin{array}{cc}
                   M  & 0 \\
                  0 & M
                \end{array} \right)\ ,\;\;\;
M = \left( \begin{array}{ccc}
                m_u & 0 & 0 \\
                0 & m_d & 0 \\
                0 & 0 & m_s
                \end{array} \right),
\end{eqnarray*}
where $m_u$, $m_d$ and $m_s$ are the quark masses.

The special role of the $\eta'$ in the quenched approximation becomes clear
from considering the quadratic part of $L$ for the $\eta'$ and ${\tilde\eta}'$
fields which (for degenerate quark masses) reads
\begin{eqnarray}
L_{\rm quad} \sim
&\left(\begin{array}{cc}\eta' & {\tilde\eta}'\end{array}\right)
\Biggl[(p^2-m_\pi^2)
\left(\begin{array}{cc}
      1 & 0 \\
      0 & -1
      \end{array} \right) \nonumber \\
& \ -\mu^2
\left(\begin{array}{cc}
      1 & -1 \\
      -1 & 1
      \end{array} \right) \nonumber
\Biggr]
\left(\begin{array}{cc}
      \eta' \\
      {\tilde\eta}'
      \end{array} \right).
\end{eqnarray}
This leads to a propagator in the $\eta$-${\tilde\eta}'$ sector
\begin{equation}
\frac{1}{p^2-m_\pi^2}
\left(\begin{array}{cc}
      1 & 0 \\
      0 & -1
      \end{array} \right)
+\frac{\mu^2}{(p^2-m_\pi^2)^2}
\left(\begin{array}{cc}
      1 & 1 \\
      1 & 1
      \end{array} \right),
\end{equation}
where $\mu^2=\frac{1}{2}V_0''(0)$ is the parameter which in the full theory
gives the singlet part of the $\eta'$ mass.  (We have set $V_5=0$ for
simplicity.  One can show that vertices coming from $V_5$ contribute only
at higher order in a combined expansion in $1/N_c$ and $M$ to quantities
considered in this talk.)

Because $\mu^2$ appears in the numerator of eq. (1), it is clear that
the $\eta'$ does not decouple for large $\mu^2$
in quenched ChPT, unlike the case of full QCD.
This can also intuitively be understood from the ``quark flow" approach.
For a discussion and applications of the quark flow
approach, see refs. \cite{Sharpe1,Sharpe2,Sharpe3,BerGolPRD}.

\section{Comparison between quenched and full ChPT}

In full ChPT the ratio $f_K/f_\pi$ at one loop for $m_u=m_d\equiv m$ is
\cite{GasLeut}
\begin{eqnarray*}
\lefteqn{\left(\frac{f_K}{f_\pi}\right)^{\rm 1-loop}_{\rm full}
= 1+\frac{5}{4}\mu_\pi-\frac{1}{2}\mu_K-\frac{3}{4}\mu_\eta} \\
&& +(m_s-m){\tilde L}_{5,\rm full},
\end{eqnarray*}
where $\mu_P=\frac{1}{16\pi^2f^2}m_P^2\log{\frac{m_P^2}{\Lambda^2}}$ and
${\tilde L}_{5,\rm full}$ is the coefficient of an $O(p^4)$ term.  In quenched
ChPT the result is \cite{BerGolJap,BerGolPRD}
\begin{eqnarray*}
\lefteqn{\left(\frac{f_K}{f_\pi}\right)^{\rm 1-loop}_{\rm qu}
= 1+} \\
&&\frac{\mu^2/3}{16\pi^2f^2}\left[
\frac{m_K^2}{2(m_K^2-m_\pi^2)}\log{\left(\frac{2m_K^2}{m_\pi^2}-1\right)}
-1\right] \\
&& +(m_s-m){\tilde L}_{5,\rm qu}.
\end{eqnarray*}
These two expressions can however not be compared directly, since in
general the parameters in the full and quenched ChPT lagrangians will
not be equal: ${\tilde L}_{5,\rm qu}\ne
{\tilde L}_{5,\rm full}$.  This implies that in order to compare full
and quenched ChPT we will have to consider physical quantities independent
of bare parameters.  In the full theory, such a quantity is
$f_\eta f_\pi^{\frac{1}{3}}f_K^{-\frac{4}{3}}$
\cite{GasLeut}.  However, in quenched
ChPT with $m_s\ne m$, the $\eta$ inherits the $\eta'$ double pole
(eq. (1)) through mixing.

A similar physical quantity can be defined in $N_f=4$ QCD with quark masses
$m_u=m_d\equiv m$ and $m_s=m_{s'}\equiv m'$.
The quantity we wish to consider here
is
\[
R=\frac{f_K}{\sqrt{f_\pi f_{\pi'}}},
\]
where $\pi$, $K$ and $\pi'$ denote the ${\bar d}u$, ${\bar s}u$ and
${\bar s'}s$ mesons, respectively.
This quantity is natural to compute in a quenched simulation, since
one has the quark propagators for various masses.  The total number of
flavors, $N_f$, is irrelevant, as only the valence quarks play a role in
the calculation of any quenched quantity.
In full ChPT,
\begin{equation}
R_{\rm full}=1-\frac{1}{64\pi^2f^2}\left[m_\pi^2\log{\frac{m_K^2}{m_\pi^2}}
+m_{\pi'}^2\log{\frac{m_K^2}{m_{\pi'}^2}}\right],
\end{equation}
whereas in quenched ChPT,
\begin{equation}
R_{\rm qu}=1+\frac{\mu^2/3}{16\pi^2f^2}\left[
\frac{m_\pi^2+m_{\pi'}^2}{2(m_{\pi'}^2-m_\pi^2)}
\log{\frac{m_{\pi'}^2}{m_\pi^2}}-1\right],
\end{equation}
where we used the tree level relation $m_K^2=\frac{1}{2}(m_\pi^2+m_{\pi'}^2)$.
Note that the logs in the full and quenched expression have completely
different forms, stemming from their different physical origins.

To get a feeling for the numerical difference we have substituted two sets
of numerical values for the quantities appearing on the right-hand side
of eqs. (2,3).  First, we have considered the ``real" world, with
$m_\pi=140$~MeV and $m_K=494$~MeV (with $m_{\pi'}$ computed from its tree
level relation to $m_\pi$ and $m_K$), and with $\mu^2/3=(500\ {\rm MeV})^2$
from the full QCD $\eta'$ mass.  Secondly, we have used the
quark and meson masses from a typical simulation
(ref. \cite{Gupetal}, $\beta=6.2$),
in particular $m=0.007$, $m'=0.03$ and $a^{-1}=
2.9$~GeV.  (Again we took $\mu^2/3=(500\ {\rm MeV})^2$.)
For the ``real" world we get
\[
R_{\rm full}=1.023,\;\;\;R_{\rm qu}=1.066
\]
(a difference of $4.2 \%$), whereas for the $\beta=6.2$ data we find
\[
R_{\rm full}=1.022,\;\;\;R_{\rm qu}=1.014
\]
(a difference of $0.8 \%$).

We have also looked at an $N_f=3$ quantity independent of the bare parameters
(using results from \cite{BerGolPRD}),
namely
\[
\chi=\frac{\langle{\bar d}d\rangle}{\langle{\bar u}u\rangle}
-\frac{m_{K^0}^2-m_{K^+}^2}{m_{K^0}^2-m_{\pi^+}^2}
\frac{\langle{\bar s}s\rangle}{\langle{\bar u}u\rangle}.
\]
(In QCD, the divergences of $\langle{\overline q}q\rangle$ cancel in $\chi$
\cite{GasLeut}.)
At tree level, the values for full and quenched ChPT coincide, but again,
at one loop
\[
\Delta\chi\equiv\chi^{\rm 1-loop}-\chi^{\rm tree}
\]
differs.

For the real world, we find
\[
\chi^{\rm tree}=0.982
\]
and
\[
\Delta\chi_{\rm full}=-0.002,\;\;\;\Delta\chi_{\rm qu}=-0.064
\]
(the difference between $\chi_{\rm full}$ and $\chi_{\rm qu}$ is
$6.3 \%$).  From the $\beta=6.2$ data of ref. \cite{Gupetal} with quark masses
$m_u=0.01$, $m_d=0.02$ and $m_s=0.03$ we obtain
\[
\chi_{\rm tree}=\frac{1}{2},
\]
\[
\Delta\chi_{\rm full}=-0.017,\;\;\;\Delta\chi_{\rm qu}=-0.026
\]
(a difference of $1.9 \%$ in $\chi$).

The one-loop corrections for
the full and quenched quantities are unrelated for both $R$ and $\chi$
(for an example where they are the same, see ref. \cite{Sharpe2}).
However,
for both examples ($R$ and $\chi$), the difference between the full and
quenched case is small.  This is due simply to the fact that the one-loop
corrections themselves are small. Note that
the differences between the quenched and full case
grow with decreasing quark mass, due to the fact that
the quenched one-loop corrections grow with decreasing quark mass.
We will return to this point in the next section.
To our knowledge,
the results presented in this section constitute  the first controlled
calculation of the difference between quantities in the quenched and full
theories.

We have been using an estimate for $\mu^2$ from the full theory, whereas
of course its quenched value need not be equal to the full value.  It
would therefore be nice if we could measure $\mu^2_{\rm qu}$ from quenched
data.  One could in principle determine $\mu^2_{\rm qu}$ from a fit of
the pion mass to the form \cite{BerGolPRD}
\[
(m_\pi^2)^{\rm 1-loop}_{\rm qu}=
Am\left[1-\frac{\mu^2/3}{8\pi^2f^2}\log{\frac{m}{m_0}}+Bm\right].
\]
We have tried to fit both the $\beta=6.2$ and $\beta=6.0$ data from
ref. \cite{Gupetal} (not only for $m_\pi$ but also for
$\langle{\bar\psi}\psi\rangle$).  However, we have not been able to uncover
any chiral logs in these data; one apparently needs smaller errors and
lower quark masses.

\section{Problems with quenched QCD?}

Let us again consider the quenched result for $f_K/f_\pi$ (cf. sect. 2).
Rewritten in
terms of quark masses the result is
\begin{eqnarray*}
\lefteqn{\left(\frac{f_K}{f_\pi}\right)^{\rm 1-loop}_{\rm qu}
=} \\
&& 1+\frac{\mu^2/3}{16\pi^2f^2}\left[
\frac{m_u+m_s}{2(m_s-m_u)}\log{\frac{m_s}{m_u}}
-1\right] \\
&& +L-{\rm term}.
\end{eqnarray*}
It is clear that the limit $m_u\to 0$
with $m_s$ fixed, which is well defined in the full theory, does not exist
here \cite{BerGolJap,BerGolPRD}.  Stated
in other words, taking both $m_u,\;m_s\to 0$, one is
left with a result dependent on the ratio $m_s/m_u$.
(A similar problem in $\langle{\overline\psi}\psi\rangle$ has been discussed
in ref. \cite{Sharpe2}.  See also ref. \cite{Sharpe3}.)
This seems to imply
that {\it no chiral limit exists for quenched QCD!}  We note here that the
bare quark mass appearing in the chiral lagrangian is analytic in the
bare quark mass which appears in the QCD lagrangian (in both the full
and the quenched
cases).  This means that redefinitions of the quark mass \cite{Sharpe2}
cannot solve the problem.

The problem arises from IR divergences due to the double
pole term in the $\eta'$ propagator (cf. eq. (1)).  The first question one may
therefore ask is whether this double pole term is really there, or whether
it might be softened by summing the quenched perturbative expansion.

This can be phrased as follows (for simplicity we consider the case of
degenerate quark masses).  In the summed theory, the parameter
$\mu_{\rm qu}^2$ appearing in eq. (1) will become momentum dependent.
(In the quenched theory
we may take eq. (1) as a definition of $\mu_{\rm qu}^2(p)$ to all orders.
Similarly, in the full theory $\mu_{\rm full}^2(p)$ is defined  to all orders
as the singlet part of the $\eta'$ self-energy.)
If now $\mu_{\rm qu}^2(p)\to 0$
for $p\to 0$, the double pole would be softened, leading to a less
divergent chiral limit.  It appears however, that this will not
occur:

We first observe that to leading order in $1/N_c$
\[
\mu_{\rm qu}^2(p)=\mu_{\rm full}^2(p)\left(1+
O\left(\frac{1}{N_c}\right)\right),
\]
and secondly, that
\[
\mu_{\rm full}^2(p=0)\ne 0
\]
in particular when the quark mass $m$ is set to zero.  The latter statement
follows from the fact
that the $\eta'$ is a well-behaved meson in full QCD with a nonvanishing
mass in the chiral limit.  Assuming the validity of the $1/N_c$ expansion,
we conclude that
\[
\mu_{\rm qu}^2(p=0)\ne 0,
\]
which means that the double pole is a true feature of the quenched theory!

This argument does not prove that no resummation could ameliorate the problem.
On this issue, however, we have the following remarks:
\begin{description}
\item[$\bullet$]
Sharpe has summed a class of diagrams in the case of degenerate quark masses
\cite{Sharpe2,Sharpe3}; this did {\it not} lead to a less singular result.

\item[$\bullet$]
In the
nondegenerate case there are many more relevant diagrams, and it is not
clear that a systematic resummation can be carried out.

\item[$\bullet$]
Any resummation method would have to work for each quantity that does not
have a chiral limit at one loop.  Many such quantities exist.  If
some resummation works for each quantity, one would expect that there is
a
general explanation of {\it why} it works.  We have shown above that this
explanation cannot be a softening of the $\eta'$ double pole term, which
nevertheless
seems to be the origin of the problem.
\end{description}

In view of these remarks, we believe that the nonexistence of a chiral
limit is a real problem of the quenched theory, and not an artifact of
chiral perturbation theory.

\section{Conclusions}
We have calculated in a controlled way the difference between the quenched
and full versions
of QCD for two physical quantities.  Such calculations
can be extended straightforwardly to many other physical quantities.

We have also pointed out that quenched QCD does not seem to have a
well-behaved chiral limit, due to the peculiar role of the $\eta'$ in the
quenched approximation.  We believe that this is a real problem of the
quenched theory, and not an artifact of ChPT.  In order for the quenched
approximation
to remain a key tool in lattice QCD, it will be essential to
gain a better understanding of this problem.

\section*{Acknowledgements}
We would like to thank Steve Sharpe for useful discussions.
Part of this work was carried out at Los Alamos National Laboratory and
UC Santa Barbara.
M.G. would
like to thank Rajan Gupta and the Theory Division of LANL,
and both of us would like
to thank the UCSB Physics Department,
and in particular Bob Sugar, for hospitality.
This work is supported in part by the
Department of Energy under grant number DE-2FG02-91ER40628.

\end{document}